\documentclass[prx, notitlepage, twocolumn,nofootinbib,superscriptaddress,longbibliography]{revtex4-1}
\usepackage[colorlinks=true,citecolor=blue,linkcolor=blue]{hyperref}
\usepackage{graphicx}
\usepackage{color}
\usepackage{dcolumn}
\usepackage{bm}
\usepackage{mathrsfs}
\usepackage{amsmath}
\usepackage{amssymb}
\usepackage{amsthm}
\usepackage{amsfonts}
\usepackage{enumerate}
\usepackage{latexsym}
\usepackage{hyperref}
\usepackage{centernot}
\usepackage{multirow}
\usepackage{subfigure}

\begin{document}

\title{Electronic and magnetic properties of the Jahn-Teller active fluoride $%
\mathrm{NaCrF_{3}}$ from first-principles calculations}
\author{Jianghan Bao}
\affiliation{National Laboratory of Solid State Microstructures and School of Physics,
Nanjing University, Nanjing 210093, China}
\affiliation{Collaborative Innovation Center of Advanced Microstructures, Nanjing
University, Nanjing 210093, China}
\author{Di Wang}
\thanks{Corresponding author: diwang0214@nju.edu.cn}
\affiliation{National Laboratory of Solid State Microstructures and School of Physics,
Nanjing University, Nanjing 210093, China}
\affiliation{Collaborative Innovation Center of Advanced Microstructures, Nanjing
University, Nanjing 210093, China}
\author{Hai-Zhou Lu}
\affiliation{Shenzhen Institute for Quantum Science and Engineering and Department of
Physics, Southern University of Science and Technology, Shenzhen 518055,
China}
\author{Xiangang Wan}
\affiliation{National Laboratory of Solid State Microstructures and School of Physics,
Nanjing University, Nanjing 210093, China}
\affiliation{Collaborative Innovation Center of Advanced Microstructures, Nanjing
University, Nanjing 210093, China}
\date{\today}

\begin{abstract}
In perovskite-type compounds, the interplay of cooperative Jahn-Teller
effect, electronic correlations and orbital degree of freedom leads to
intriguing properties. $\mathrm{NaCrF_{3}}$ is a newly synthesized
Jahn-Teller active fluoroperovskite where the $\mathrm{CrF_{6}^{4-}}$
octahedrons are considerably distorted. Based on the first-principles
calculation, we analyze its electronic structure and magnetic properties.
Our numerical results show that the $\mathrm{Cr^{2+}}$ ions adopt the
high-spin $t_{2g\uparrow}^{3}e_{g\uparrow}^{1}$ configuration with $G$-type orbital ordering. We also estimate the
magnetic exchange couplings and find that the in-plane and interplanar
nearest-neighbor interactions are ferromagnetic and antiferromagnetic,
respectively. The ground state of this material is $A$-type
antiferromagnetic, in agreement with the experiments. Reasonable Curie-Weiss
and $\mathrm{N\acute{e}el}$ temperatures compared to the experiments are
given by mean-field approximation theory. Our results give a complete
explanation of its electronic structure, magnetic and orbital order, and
help to further comprehend the behaviors of Jahn-Teller active
perovskite-type fluoride.
\end{abstract}

\maketitle

\section{Introduction}

In strongly correlated electron systems, the interplay between charge, spin,
orbital and lattice degrees of freedom gives rise to profuse and exotic
physics properties \cite{Imada98rmp}. Especially, the orbital physics is
usually significant and inevitable for understanding this complexity \cite%
{NagaosaN00science,KhaliullinG05ptps}. A fascinating example is the
cooperative Jahn-Teller (JT) effect, which refers to a symmetry-lowering
structural deformation driven by the coupling between the degenerate orbital
states and the collective lattice vibrations \cite%
{JahnH38prs,Gehring75rpp,Kugel82spu,KhomskiiDI03prb,Bersuker_2006}.
Accompanied with strong electronic correlations and orbital degree of
freedom, the cooperative JT effect gives rise to intriguing behaviors in
perovskite-type compounds \cite{Kugel82spu,tmo}. The best known example is
the parent material of colossal magnetoresistance (CMR) manganites, $\mathrm{%
LaMnO_{3}}$ \cite{JinS94science,SalamonMB01rmp}, where the cooperative JT
effect plays a fundamental role in stabilizing the $A$-type
antiferromagnetic (AFM) spin and $C$-type orbital ordering (OO) \cite%
{SarmaDD95prl,Solovyev96prl,PavariniE10prl,Popovic02prl}. Moreover, it
arouses the formation of polaron which is regarded as a key mechanism in CMR
effect \cite{MillisAJ95prl,MillisAJ96prl,JoossC07pnas}.

Besides in the perovskite oxides, the cooperative JT effect also causes
various remarkable magnetic and structural effects in fluoroperovskites \cite%
{HirakawaK70ptps,LiechtensteinAI95prb,MedvedevaJE02prb,TowlerM95prb,Pavarini08prl,Moreira04ijqc,LeonovI08prl,TongJ10sss,Kurzydlowski17prb,Margadonna06jacs,Margadonna07jmc,XiaoY10prb,Giovannetti08prb,XuYH08jcp,WangGT11prb,AutieriC14prb,MingX14cpb}%
. In the paradigmatic example, tetragonal $\mathrm{KCuF_{3}}$, two distinct
types of collective JT distortions lead to two isoenergetic structural
polytypes \cite%
{HirakawaK70ptps,LiechtensteinAI95prb,MedvedevaJE02prb,TowlerM95prb}. The
cooperative JT effect is not only essential for stabilizing the
antiferro-orbital order in $\mathrm{KCuF_{3}}$ at high temperature \cite%
{Pavarini08prl}, but also makes this compound one of the most ideal
one-dimensional antiferromagnets with $\mathrm{S=\frac{1}{2}}$ \cite%
{HirakawaK70ptps,tmo}. Similar uniform AFM chains induced by cooperative JT
effect are also found in triclinic $\mathrm{AgCuF_{3}}$ and $\mathrm{%
NaCuF_{3}}$ \cite{TongJ10sss,Kurzydlowski17prb}. Additionally, chromium
fluoroperovskites are also able to activate cooperative JT effect, such as $%
\mathrm{KCrF_{3}}$ \cite{Margadonna06jacs,Margadonna07jmc,XiaoY10prb}. This
material has rich structural phase-transitions ($Pm3m\rightarrow I4/mcm$ at
973K and $I4/mcm\rightarrow I112/m$ at 250k) \cite%
{Margadonna06jacs,Margadonna07jmc,XiaoY10prb} and exhibits an ordering of
staggered $3d_{3x^{2}-r^{2}}$ and $3d_{3y^{2}-r^{2}}$ orbitals in $ab$-plane
which is rotated by $90^{\circ }$ in consecutive layers along the $c$%
-direction at room temperature \cite{Margadonna06jacs,Margadonna07jmc}. The
ample structural and magnetic behaviors of $\mathrm{KCrF_{3}}$ have
attracted much interest in theoretical study in recent years \cite%
{Giovannetti08prb,MingX14cpb,XuYH08jcp,WangGT11prb,AutieriC14prb}.

Recently, another chromium fluoroperovskite $\mathrm{NaCrF_{3}}$
was successfully synthesized with a novel wet chemistry method \cite%
{BernalF20arxiv}. Incorporating the JT active Cr ions, this material
exhibits obvious cooperative structural distortions. Neutron powder
diffraction experiments revealed that $\mathrm{NaCrF_{3}}$ adopts a canted $A
$-type AFM ground state at low temperature \cite{BernalF20arxiv}. The Curie-Weiss
temperature and $\mathrm{N\acute{e}el}$ temperature were given by -4K and
21.3K, respectively \cite{BernalF20arxiv}. Its $\mathrm{N\acute{e}el}$
temperature is much lower than $\mathrm{KCrF_{3}}$ (79.5 K) \cite{XiaoY10prb}%
, indicating its weak antiferromagnet nature. Isostructural with triclinic $%
\mathrm{NaCuF_{3}}$ \cite{BernalF20arxiv,TongJ10sss}, $\mathrm{NaCrF_{3}}$
has considerably lower crystal symmetry than $\mathrm{KCrF_{3}}$. Meanwhile,
the Cr ions of $\mathrm{NaCrF_{3}}$ have $3d^{4}$ electronic configuration,
which are different from the $3d^{9}$ Cu ions in $\mathrm{KCuF_{3}}$. Thus,
it will be worthwhile to investigate the electronic structure and magnetic
properties of this distinct compound.
\begin{figure}[b]
\begin{minipage}[c]{0.25\textwidth}
			\subfigure[]{
				\includegraphics[width=1\textwidth]{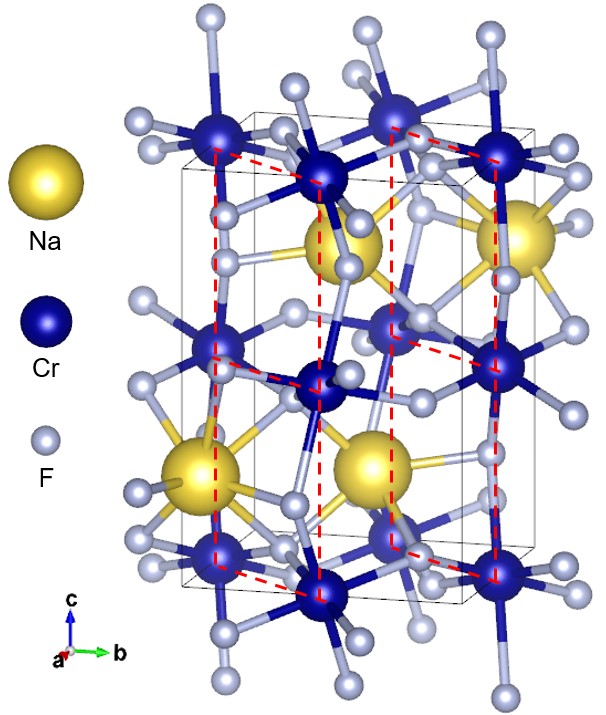}
				\label{struct1}
			}
\end{minipage}
\begin{minipage}[c]{0.2\textwidth}
			\subfigure[]{
				\includegraphics[width=1\textwidth]{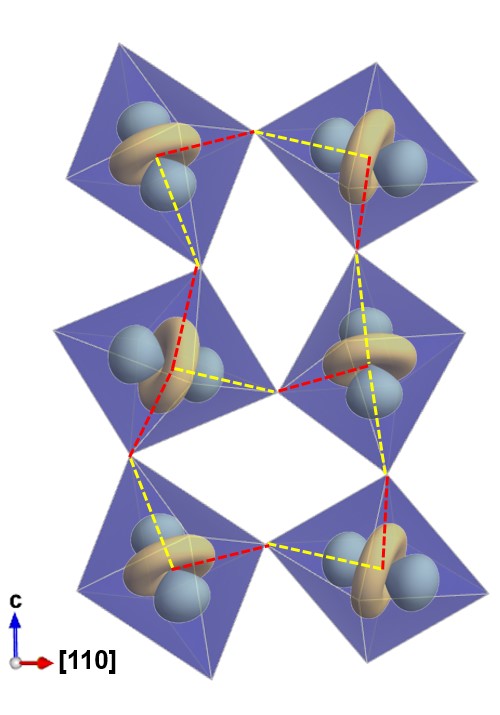}
				\label{octahedra}
			}
\end{minipage}
\caption{(a) Crystal structure of $\mathrm{NaCrF_{3}}$ \protect\cite%
{BernalF20arxiv}. Na, Cr and F atoms are represented here by yellow, blue
and grey spheres, respectively. Red dashed lines indicate the (1$\bar{1}$%
0)-plane. (b) Schematic view of occupied $d_{3z^{2}-r^{2}}$ orbitals of $%
\mathrm{Cr^{2+}}$ and corresponding distorted $\mathrm{CrF_{6}^{4-}}$
octahedrons in (1$\bar{1}$0)-plane, yellow and red dashed lines denote the
alternating long and short Cr-F bonds. Notice that the lobes of $%
d_{3z^{2}-r^{2}}$ point into the direction of long Cr-F bonds.}
\end{figure}
\begin{figure*}[tbh]
\begin{minipage}[c]{0.345\textwidth}
			\subfigure[]{
				\includegraphics[width=1\textwidth]{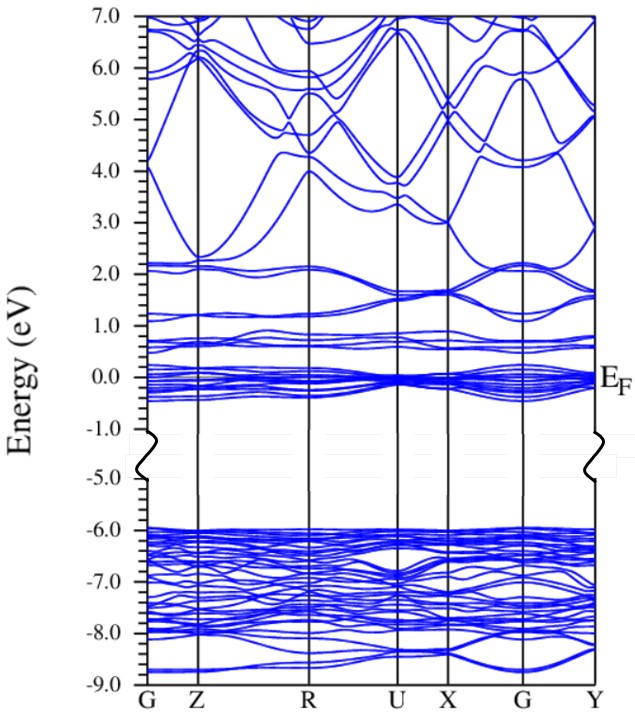}
				\label{lda}
			}
		\end{minipage}
\begin{minipage}[c]{0.3025\textwidth}
			\subfigure[]{
				\includegraphics[width=1\textwidth]{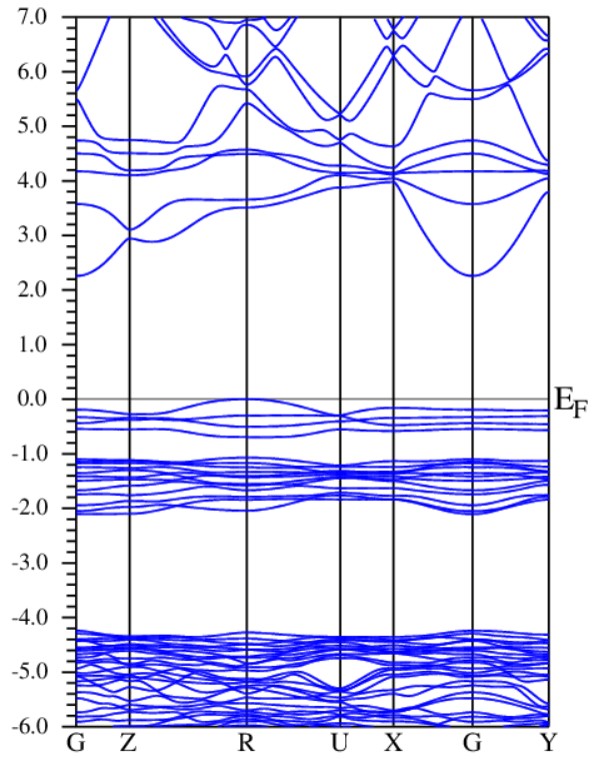}
				\label{lsdaup}
			}
		\end{minipage}
\begin{minipage}[c]{0.3025\textwidth}
			\subfigure[]{
				\centering
				\includegraphics[width=1\textwidth]{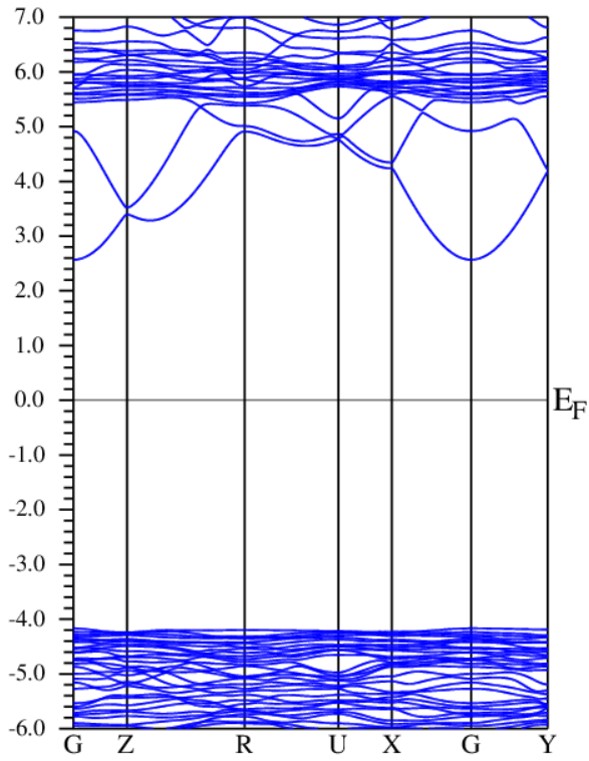}
				\label{lsdadn}
			}
		\end{minipage}		
\caption{Electronic structure of $\mathrm{NaCrF_{3}}$. (a): the
results of LDA calculation. No band occurs in the omitted part (from -5 eV
to -1 eV). (b): the spin-up channel. (c): the spin-down
channel given by LSDA + $U$ ($U$ = 6 eV) calculation with FM configuration.
The Fermi level is set to be zero.}
\end{figure*}

In this paper we systematically analyze the electronic and magnetic
properties of $\mathrm{NaCrF_{3}}$ by using the first-principles
calculation. Band structure and partial density of states (PDOS) show
clearly the splitting of $\mathrm{e_{g}}$ orbitals of $\mathrm{Cr^{2+}}$
ions caused by the axial elongation of $\mathrm{CrF_{6}^{4-}}$ octahedrons.
Our numerical results show that this material exhibits $G$-type
antiferro-orbital ordering while the $\mathrm{Cr^{2+}}$ ions adopt the
high-spin $t_{2g\uparrow }^{3}e_{g\uparrow }^{1}$ configuration. $A$-type
AFM spin state is found to own the lowest total energy in accordance with
the experiments. Based on energy-mapping procedure, a ferromagnetic (FM)
in-plane nearest-neighbor (NN) coupling, an AFM interplanar NN interaction
and a non-negligible interplanar next-nearest-neighbor (NNN) exchange
parameter are estimated. Through mean-field approximation theory, we
calculate the Curie-Weiss and $\mathrm{N\acute{e}el}$ temperatures,
which are in reasonable agreement with the results of experiments. The
ground magnetic state of $A$-type magnetic order can be understood
based on the electronic configuration of $\mathrm{Cr^{2+}}$ ions and
the Cr-F-Cr superexchange pathways.

\section{Method and crystal structure}

The density functional calculations have been performed by utilizing the full
potential linearized augmented plane wave (LAPW) method as implemented in
Wien2k code \cite{wien2k}. Local spin density approximation (LSDA) is
adopted as the exchange-correlation potential \cite{VoskoSH80cjp}. To take
into account the Coulomb repulsion of the Cr-3d electrons, LSDA + $U$ scheme
is also performed \cite{AnisimovVI97jpcm}. The value of $U_{eff}$ ($%
U_{eff}=U$-$J$, the Hund exchange parameter $J$ is set to be 0), which
varies from 4 eV to 8 eV, has been widely used in previous first-principle
calculation \cite{Giovannetti08prb,LiechtensteinAI95prb}. We found that our
numerical essential properties do not depend on the value of $U_{eff}$ in
this range and the results of $U_{eff}$ = 6 eV are mainly presented here.
The muffin-tin sphere radii are chosen to be 2.06, 1.81 and 1.90 Bohr for
Na, F and Cr atoms, respectively. The plane wave cut-off $K_{max}$ is
determined by $R_{min}K_{max}=7$, where the $R_{min}$ is the smallest of all
atomic sphere radii. The convergence criterion of the crystal total energy
is set to be 0.01 mRy per conventional unit cell, and a sufficient large k
mesh is used for the integration over the Brouillon zone. In addition to the FM
configuration, five possible AFM configurations (shown in Fig. \ref{afm})
have also been taken into consideration to explore the magnetic properties
of $\mathrm{NaCrF_{3}}$.

The experimental crystal structure (Fig. \ref{struct1}) obtained by
high-resolution synchrotron X-ray and neutron powder diffraction is used in
the calculation \cite{BernalF20arxiv}. $\mathrm{NaCrF_{3}}$ belongs to the
triclinic crystal system (space group $P\bar{1}$), with lattice constants
a=5.48428(11) \AA , b=5.67072(11) \AA\ and c=8.13620(15) \AA . The angles
between the lattice vectors are $\alpha$=$90.3860(10)^{\circ}$, $\beta$=$%
90.2816(8)^{\circ}$ and $\gamma$=$86.3255(8)^{\circ}$ \cite{BernalF20arxiv}.
Four inequivalent crystallographic sites are occupied by the Cr cations
(Fig. \ref{struct1}), which form distorted corner-sharing $\mathrm{CrF_{6}}$
octahedrons enclosed in Na cage with the surrounding $\mathrm{F}$ anions.
Each axially elongated $\mathrm{CrF_{6}}$ octahedra has 2 long and 4 short
Cr-F bonds. It should also be noted that, the $\mathrm{CrF_{6}}$ octahedrons
connect with each other through alternating long-short Cr-F bonds within the
(1$\bar{1}$0) plane, while all the interplanar connection consists of short
bonds (Fig. \ref{octahedra}).

\section{Results and discussion}

\begin{figure}[bh]
	\centering
	\begin{minipage}[c]{0.45\textwidth}
		\subfigure{
			\includegraphics[width=1\textwidth]{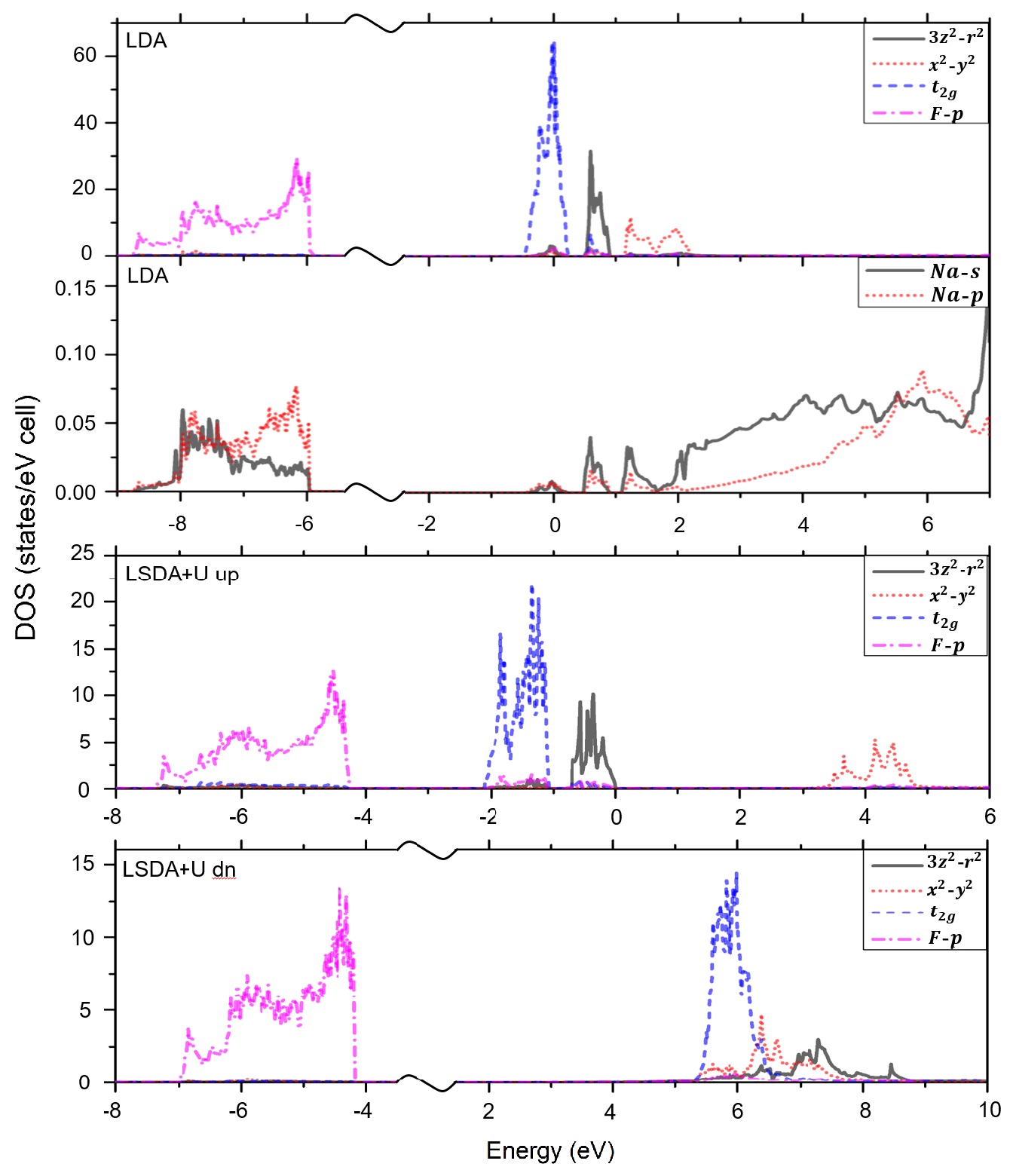}
			\label{octahedra}
		}
	\end{minipage}
	\caption{Partial density of states (PDOS) of $\mathrm{NaCrF_{3}}$ given by the LDA
		and LSDA + $U$ ($U$ = 6 eV) calculations. The local coordinate system is used with the $z$-direction defined along the longest Cr-F bond of each $\mathrm{CrF_{6}}$ octahedron in the calculation of PDOS.}
	\label{dos}
\end{figure}

LDA calculation is carried out firstly to clarify the basic electronic
features. The band structures and the density of states (DOS) are given in
Fig. \ref{lda} and \ref{dos}. The $z$-direction of local coordinate
system is defined along the longest Cr-F bond of each $\mathrm{CrF_{6}}$
octahedron in the analysis of PDOS. There are 12 F ions in the unit cell. As shown in Fig. \ref{lda} \& \ref{dos}, 36 F-2$p$ bands mainly locate in the
energy range from -9 eV to -6 eV, indicating that the F-2$p$ orbitals are nearly
full occupied. In comparison, the F-2$p$ bands of $\mathrm{KCuF_{3}}$, which
locates between -7 eV and -3 eV, are closer to the Fermi level due to the
stronger electronegativity of Cu than Cr. As shown in Fig. \ref{dos}, the
main contribution of Na bands is above the Fermi level, while there is also
small distribution of Na states between -9 and -6 eV, indicating the
non-negligible hybridization between Na and F states. Thus the nominal
chemical valence of F and Na can be regarded as -1 and +1, respectively. As
a consequence, the nominal valence of Cr is +2 and the outer shell
electronic configuration of Cr ion is $3d^{4}$. Our results also show that
the bands around the Fermi level consists mostly of the Cr-3$d$ electrons
(Fig. \ref{dos}). In each axially elongated $\mathrm{CrF_{6}}$ octahedron,
two long Cr-F bonds are collinear, while the angles between the long bonds
and the short bonds deviate slightly from the right angle (ranging from $%
84.50^{\circ }$ to $96.17^{\circ }$) \cite{BernalF20arxiv}. Though the $\mathrm{CrF_{6}}$ octahedrons deform from the ideal ones, the Cr-3$d$ orbitals are still roughly divided into 12 $t_{2g}$ bands (from -0.46 eV to 0.25 eV) and 8\ $e_{g}$ bands (from 0.46 eV to 2.53 eV), as shown in Fig. \ref{lda} \& \ref{dos}. Furthermore, influenced by the axial elongation of the $\mathrm{CrF_{6}}$
octahedrons, the $e_{g}$ states split into 4 $d_{3z^{2}-r^{2}}$ bands from
0.46 eV to 0.91 eV and 4 $d_{x^{2}-y^{2}}$ bands from 1.08 eV to 2.53 eV,
as shown in Fig. \ref{lda} and \ref{dos}. The energy\ splitting between the
band centers of $d_{3z^{2}-r^{2}}$ and $d_{x^{2}-y^{2}}$ states is estimated
to be about 1.05 eV.

As shown in Fig. \ref{dos}, the DOS at Fermi level is quite large, which
indicates the electronic instability. To explore its magnetic properties,
the LSDA + $U$ calculation of FM configuration is performed. The band structures
are shown in Fig. \ref{lsdaup}\ \&\ \ref{lsdadn}. The spin exchange splitting is relatively large and the down-spin bands lie far away from the Fermi level.
As shown in the band structure and DOS (Fig. \ref{lsdaup}\ \&\ \ref{lsdadn} \&\ \ref{dos}), the $t_{2g}$ and $d_{3z^{2}-r^{2}}$ orbitals in spin-up channel are fully
occupied, while the $d_{x^{2}-y^{2}}$ orbitals are entirely empty in the FM
structure. Thus, the Cr $3d^{4}$ configuration can be regarded as holding a $%
t_{2g\uparrow }^{3}e_{g\uparrow }^{1}$ high-spin state. Enhanced by the
combination of JT effect and electronic correlation, the energy\ splitting
between the weight-centers of $d_{x^{2}-y^{2}}$ and $d_{3z^{2}-r^{2}}$ states
in spin-up channel is estimated to be 4.5 eV. Note that the occupied $%
d_{3z^{2}-r^{2}}$ orbitals form a $G$-type orbital ordering pattern, as
shown in Fig. \ref{octahedra}. The high-spin state $\mathrm{Cr^{2+}}$ ions
and $G$-type\ orbital order show remarkable similarity with those of $%
\mathrm{KCrF_{3}}$ \cite{Giovannetti08prb,MingX14cpb,XuYH08jcp,WangGT11prb}.
\begin{figure}[tbh]
\centering
\subfigure[]{
			\includegraphics[width=0.147336\textwidth]{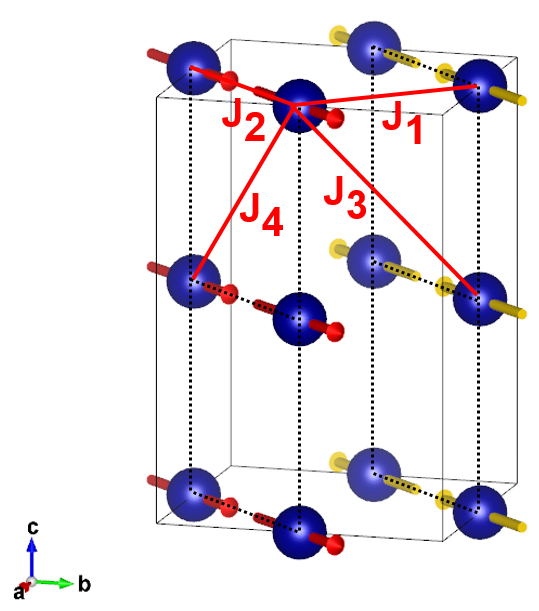}
			\label{afm1}
		}
\subfigure[]{
			\includegraphics[width=0.12\textwidth]{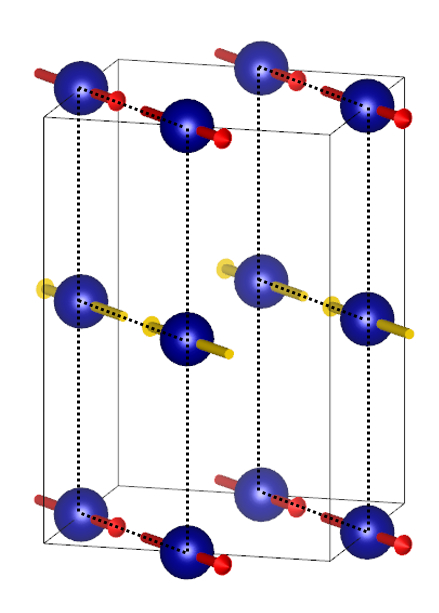}
			\label{afm2}
		}
\subfigure[]{
			\includegraphics[width=0.12\textwidth]{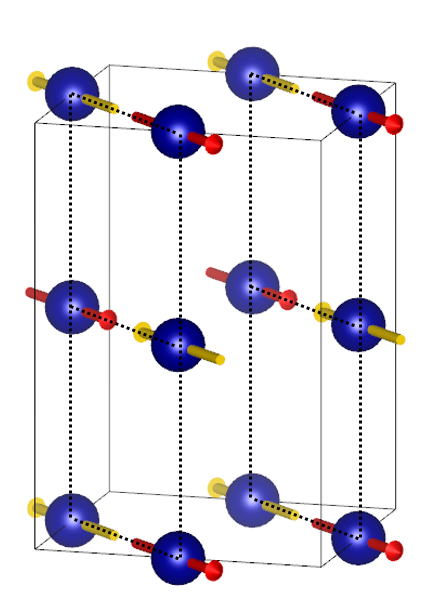}
			\label{afm3}
		}
\subfigure[]{
			\includegraphics[width=0.12\textwidth]{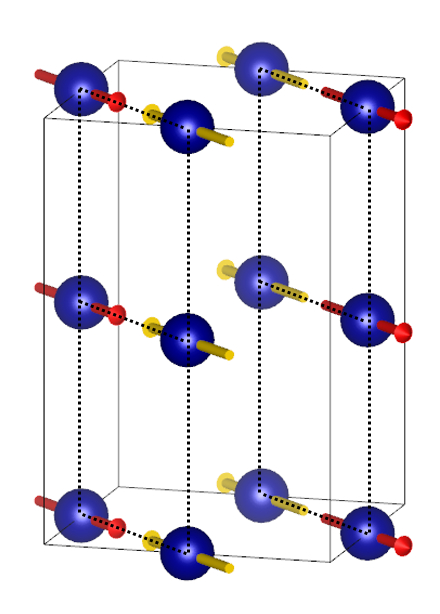}
			\label{afm4}
		}
\subfigure[]{
			\includegraphics[width=0.12\textwidth]{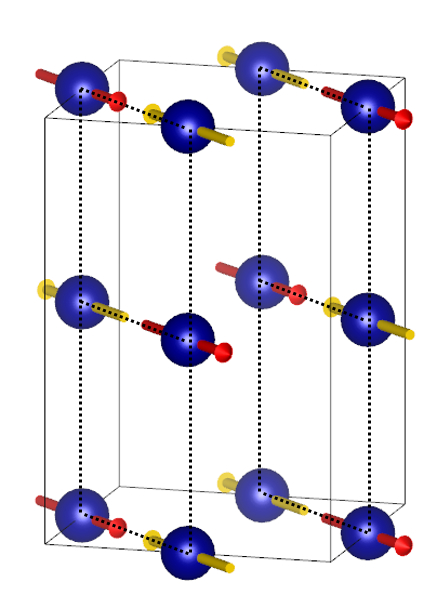}
			\label{afm5}
		}
\caption{Five AFM configurations considered in our DFT calculations. (a),
(b), (c), (d) and (e) represent AFM-1, AFM-2, AFM-3, AFM-4 and AFM-5 spin
structures respectively. Blue spheres stand for the Cr atoms within the $(1%
\bar{1}0)$-plane which is represented by the dashed lines. Red arrows
denotes the spins in up-direction and yellow arrows denotes the spins in
down-direction. The 4 exchange parameters that we take into consideration
are shown in (a).}
\label{afm}
\end{figure}

In order to study the magnetic ground state of $\mathrm{NaCrF_{3}}$, five
AFM configurations have been taken into consideration in addition to the FM
state (See Fig. \ref{afm} for the schematic descriptions of the different
AFM configurations). AFM-1 and AFM-2 are two different $A$-type AFM
structures where the AFM couplings are in [1$\bar{1}$0]- and
[001]-direction, respectively. The $\mathrm{Cr^{2+}}$ ions
of AFM-3 configuration couple anti-ferromagnetically in the $(1\bar{1}0)$-plane, but
ferromagnetically in [1$\bar{1}$0]-direction. And in AFM-4, the magnetic moments are anti-parallel in $(001)$-plane, but parallel in [001]-direction. In the spin configuration
AFM-5, each spin is set to be anti-parallel to all its nearest neighbors.
Our numerical results reveal that different magnetic configurations have
minor influence on the electronic configuration and the orbital ordering of this material, and all magnetic configurations have the same high-spin $t_{2g\uparrow }^{3}e_{g\uparrow }^{1}$ state and the $G$-type orbital order. The differences
between total energies and magnetic moments of $\mathrm{Cr^{2+}}$ of various magnetically ordered structures for LSDA + $U$ calculations are summarized in Table \ref{energy}. The results show that the AFM-1 configuration has the lowest total energies under different $U$ values, which is in agreement with the experimental
results \cite{BernalF20arxiv}. The calculated magnetic moments of
different spin configurations are similar, which indicates that the
magnetism of $\mathrm{NaCrF_{3}}$ is quite localized and the total energy
differences mainly originates from the inter-atomic exchange interactions.
Thus, it allows us to perform the energy-mapping procedure to estimate the
exchange couplings.

The exchange couplings are analyzed by using the Heisenberg Hamiltonian:
\begin{equation}
H=\sum_{i<j}J_{ij}\mathbf{S}_{i}\cdot\mathbf{S}_{j}  \label{heisenberg}
\end{equation}
where $J_{ij}$ stands for the spin exchange parameter between two spins at
sites $i$ and $j$. We have calculated the exchange parameters $J_{ij}$ by
using the relative total energies of different magnetic ordering systems.
The exchange interactions that we consider are shown in Fig. \ref{afm1},
where $J_{1}$ and $J_{3}$ denote the NN and NNN interactions between the (1$%
\bar{1}$0)-planes, while $J_{2}$ and $J_{4}$ represent the NN and NNN
couplings within the (1$\bar{1}$0)-planes.
\begin{table}[ht]
\centering
\begin{tabular}[t]{cc|cccccc}
\hline\hline
&  & FM & AFM-1 & AFM-2 & AFM-3 & AFM-4 & AFM-5 \\ \hline
\multirow{3}{*}{$E_{tot}$} & $U$=4 eV & 5.15 & 0 & 8.53 & 9.29 & 7.35 & 10.34
\\
& $U$=6 eV & 4.85 & 0 & 6.65 & 5.84 & 5.21 & 7.38 \\
& $U$=8 eV & 4.52 & 0 & 5.69 & 4.00 & 4.34 & 5.67 \\ \hline
\multirow{3}{*}{$M_{Cr}$} & $U$=4 eV & 3.366 & 3.361 & 3.361 & 3.356 & 3.356 &
3.351 \\
& $U$=6 eV & 3.397 & 3.394 & 3.394 & 3.390 & 3.388 & 3.387 \\
& $U$=8 eV & 3.425 & 3.423 & 3.423 & 3.420 & 3.419 & 3.415 \\ \hline\hline
\end{tabular}
\caption{Relative total energies ($E_{tot}$ in meV/f.u.) of different
magnetic ordering states and corresponding magnetic moment ($M_{Cr}$ in $%
\protect\mu_{B}$) of $\mathrm{Cr^{2+}}$, total energy of AFM-1 structure is
set to be 0.}
\label{energy}
\end{table}

By applying the spin Hamiltonian model (Eq.\ref{heisenberg}) on the 6
different magnetic configurations, the total energy per conventional unit
cell are expressed as:
\begin{eqnarray}
E_{FM}&=&E_{0}+4(J_{1}+2J_{2}+4J_{3}+2J_{4})S^{2}  \notag \\
E_{AFM1}&=&E_{0}+4(-J_{1}+2J_{2}-4J_{3}+2J_{4})S^{2}  \notag \\
E_{AFM2}&=&E_{0}+4(J_{1}-2J_{4})S^{2}  \notag \\
E_{AFM3}&=&E_{0}+4(J_{1}-2J_{2}-4J_{3}+2J_{4})S^{2}  \notag \\
E_{AFM4}&=&E_{0}+4(-J_{1}-2J_{4})S^{2}  \notag \\
E_{AFM5}&=&E_{0}+4(-J_{1}-2J_{2}+4J_{3}+2J_{4})S^{2}  \notag
\end{eqnarray}
Here $E_{0}$ denotes the paramagnetic part of the total energy, which is
considered as irrelevant of the change of spin configuration. The values of
the exchange parameters $J$ can be evaluated by mapping these energies
obtained with LSDA + $U$ calculations. Since the number of magnetic
configurations is larger than the number of exchange parameters, a least
squares method is applied.

The calculated exchange couplings from energy-mapping analysis are given in
Table \ref{jpara}. Generally speaking, the magnetic exchanges are
considerably weaker in $\mathrm{NaCrF_{3}}$ than in other materials with
similar Jahn-Teller active ions \cite%
{TongJ10sss,Giovannetti08prb,MingX14cpb,XuYH08jcp}. With the value of $U$
increasing, the values of most magnetic couplings decrease as expected. The
dominant terms are $J_{1}$ and $J_{2}$, where the positive $J_{1}$ indicates
an interplanar AFM interaction, while the negative $J_{2}$ reveals a FM
coupling within the (1$\bar{1}$0)-plane. The results correspond well with
experiments, since the combination of these NN interactions leads to the $A$%
-type magnetic ground state. As shown in Table \ref{jpara}, $J_{1}$ and $%
J_{2}$ have same order of magnitude, indicating that $\mathrm{NaCrF_{3}}$
exhibits three-dimensional magnetic nature. Moreover, we also get a
non-negligible positive interplanar NNN exchange constant $J_{3}$, which
strengthens the AFM interactions between the (1$\bar{1}$0)-planes.
\begin{table}[th]
\centering
\begin{tabular}[t]{ccccc}
\hline\hline
& \quad $J_{1}$\quad & \quad $J_{2}$\quad & \quad $J_{3}$\quad & \quad $J_{4}
$\quad \\ \hline
$U$=4 eV & \quad 0.211\quad & \quad -0.445\quad & \quad 0.101\quad & \quad
-0.035\quad \\
$U$=6 eV & \quad 0.250\quad & \quad -0.269\quad & \quad 0.096 \quad & \quad
-0.031\quad \\
$U$=8 eV & \quad 0.183\quad & \quad -0.158\quad & \quad 0.092 \quad & \quad
-0.004\quad \\ \hline\hline
\end{tabular}
\caption{Calculated exchange constants (meV) for different computational
settings.}
\label{jpara}
\end{table}

Based on the spin exchange parameters, we calculate the Curie-Weiss and $%
\mathrm{N\acute{e}el}$ temperature by the mean-field theory approximation
\cite{SmartJS66}:
\begin{eqnarray}
\theta & = & \frac{S(S+1)}{3k_{B}}\left(\sum_{i}z_{i}J_{i}+\sum_{j}z^{\prime
}_{j}J^{\prime }_{j}\right) \label{cwt} \\
T_{N} & = &\frac{S(S+1)}{3k_{B}}\left(\sum_{i}z_{i}J_{i}-\sum_{j}z^{\prime
}_{j}J^{\prime }_{j}\right) \label{nt}
\end{eqnarray}
where $J_{i}$ and $z_{i}$ are the neighbor exchange parameters between the
sites with same spin orientation and the corresponding number of adjacent
neighbors, whereas $J^{\prime }_{j}$ and $z^{\prime }_{j}$ represent the
magnetic coupling and the coefficients between the sites with opposite spin
orientation.

According to the experiments \cite{BernalF20arxiv}, Curie-Weiss temperature $%
\theta $ is -4 K while $\mathrm{N\acute{e}el}$ temperature $T_{N}$ is 21.3
K, smaller than that of $\mathrm{KCrF_{3}}$ ($T_{N}$ = 79.5 K) \cite%
{XiaoY10prb}. Since the Curie-Weiss temperature represents sum of all
magnetic couplings in the mean-field theory (Eq. \ref{cwt} \& \ref{nt}),
the relative low value of the frustration index $\frac{\theta }{T_{N}}$
(smaller than 1) reveals the interplay of the FM and AFM interactions \cite{BaralPR17jac}, in agreement with our results shown in Table \ref{jpara}. When $U$
equals 6 eV (reasonable value compared with similar material \cite{Giovannetti08prb,LiechtensteinAI95prb}), Curie-Weiss temperature is
estimated to -1.67 K while the $\mathrm{N\acute{e}el}$ temperature is
about 57.31 K. The results can be regarded as qualitatively consistent
with the experimental ones.
\begin{figure}[tbh]
\centering
\subfigure{
			\includegraphics[width=0.45\textwidth]{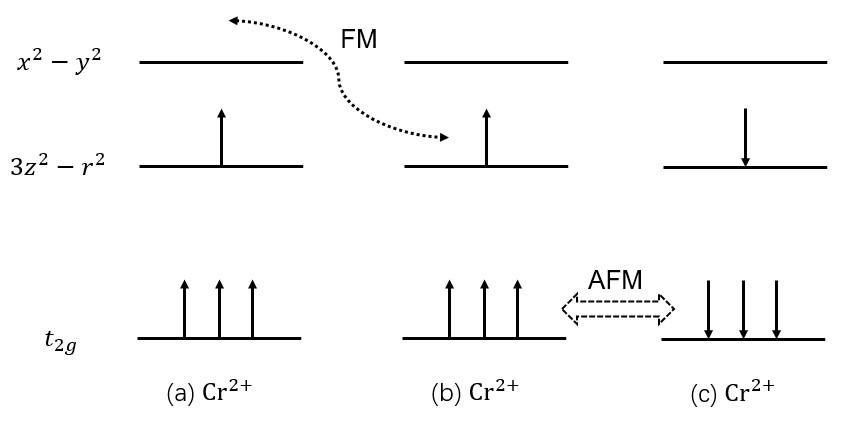}
		}
\caption{Diagram of crystal levels of high-spin $\mathrm{Cr^{2+}}$ of
different $\mathrm{Cr^{2+}}$ sites. The notations of the orbitals are given
based on the local coordinate system of each $\mathrm{Cr^{2+}}$. (a) and (b)
stand for two NN $\mathrm{Cr^{2+}}$ within the (1$\bar{1}$0)-plane, while
(c) represents the NN $\mathrm{Cr^{2+}}$ of (b) in another plane. The
in-plane $e_{g}$ electron hopping leads to FM interaction while the
interplanar coupling between $t_{2g}$ orbitals results in AFM interaction.}
\label{path}
\end{figure}

It should be noted that the ground state with $A$-type AFM structure can be
understood based on the superexchange pathways and electronic occupation (Fig. \ref{dos}) in orbital order. Based on our first-principles calculations, schematic crystal field levels and two possible exchange
pathways are presented in Fig. \ref{path}. Two NN $\mathrm{Cr^{2+}}$ ions within the (1$\bar{1}$0)-plane are given by (a) and (b), while (c) represents
the interplanar NN $\mathrm{Cr^{2+}}$ ion of (b). Note that the local $%
d_{3z^{2}-r^{2}}$ states points to the direction of longest Cr-F bond while
the pathways in the (1$\bar{1}$0)-plane connect through alternating
long-short Cr-F bonds as shown in Fig. \ref{octahedra}, therefore the
hopping between these $d_{3z^{2}-r^{2}}$ orbitals would be small.
Within the (1$\bar{1}$0)-plane, the in-plane superexchange couplings are
mainly via the hopping between the $d_{3z^{2}-r^{2}}$ and $d_{x^{2}-y^{2}}$
orbitals. According to the second Goodenough-Kanamori-Anderson (GKA) rule
\cite{GoodenoughJB63,KanamoriJ59jpcs,AndersonPW63ssp,Kugel73jetp}, a FM
interaction within the (1$\bar{1}$0)-plane\ emerges through the virtual
hopping of the half-filled $d_{3z^{2}-r^{2}}$ state and the empty $%
d_{x^{2}-y^{2}}$ state. As for the interplanar magnetic interaction,
since the $d_{3z^{2}-r^{2}}$ states are lying in the (1$\bar{1}$0)-plane
and the $d_{x^{2}-y^{2}}$ states are unoccupied, the interplanar hopping
between $e_{g}$ states is small. Thus the interplanar exchange are mainly via the virtual hopping between $t_{2g}$ orbitals. By applying the first GKA rule \cite%
{GoodenoughJB63,KanamoriJ59jpcs,AndersonPW63ssp,Kugel73jetp}, we have that
the interplanar exchange between the half-filled $t_{2g}$ states is AFM,
since the hopping between\ half-filled states with same spin orientation is
forbidden by the Pauli exclusion principle. These together make the magnetic
ground state of $\mathrm{NaCrF_{3}}$ to be the AFM-1 magnetic configuration.

\section{Conclusion}

In conclusion, we have presented comprehensive investigation of a newly
synthesized JT active ternary fluoroperovskite $\mathrm{NaCrF_{3}}$ through
the DFT calculation. The high-spin configuration of $\mathrm{Cr^{2+}}$ and
the $G$-type orbital ordering are proposed by our numerical results. We also
confirmed the $A$-type AFM magnetic ground state obtained by the experiments.
With the energy-mapping procedure, we estimate the exchange parameters. The
interplanar NN interaction $J_{1}$ is AFM while in-plane NN coupling $J_{2}$
is FM. Reasonable Curie-Weiss and $\mathrm{N\acute{e}el}$ temperatures are
also estimated by using mean-field approximation theory. Based on the
GKA rule, the magnetic ground state of $A$-type AFM order is
understood. We hope our calculation of this compound may help to further
understand of the behaviors in JT active perovskite fluoride.

\section{Acknowledgements}

This work is supported by the National Natural Science Foundation of China
(NSFC) (Grants No. 11834006, No. 11525417, No. 51721001, and No. 11790311),
National Key R\&D Program of China (Grants No. 2018YFA0305704 and No.
2017YFA0303203). H. Z. Lu is supported by the NSFC (11925402). X. Wan also
acknowledges the support from the Tencent Foundation through the XPLORER
PRIZE.

\bibliography{nacrf3}

\end{document}